\documentclass[aps,prb,twocolumn,showpacs]{revtex4}
\pdfoutput=1
\usepackage{epsfig}
\usepackage{amsmath}
\usepackage{amssymb}
\usepackage[colorlinks=true,citecolor=blue,linkcolor=blue]{hyperref}

\begin{document}

\title{Topological Insulators on the Decorated Honeycomb Lattice}
\author{Andreas R\"uegg, Jun Wen,$^*$ and Gregory A. Fiete}
\affiliation{Department of Physics, The University of Texas at Austin, Austin, Texas 78712, USA}

\date{\today}

\begin{abstract}

We show that the decorated honeycomb (``star") lattice supports a number of topological insulating phases with a non-trivial $Z_2$ invariant and time-reversal symmetry protected gapless edge modes.  We investigate the stability of these phases with respect to various symmetry breaking perturbations and demonstrate the connection to the  recent discovery of an exactly solvable $S=1/2$ chiral spin liquid model [Phys. Rev. Lett. {\bf 99}, 247203 (2007)] with non-Abelian and Abelian excitations on the same lattice at strong interaction strength.  Our work highlights the relationship between non-interacting topological band insulators and strongly interacting topologically ordered spin systems, and points to promising avenues for enlarging the number of known examples of both. 

 \end{abstract}



\maketitle
\section{Introduction}
In recent years, the study of various types of topological order in condensed matter physics has  dramatically increased.\cite{VolovikBook,WenBook, Nayak:rmp08} The interest in this topic has been driven in large part by the fractional quantum Hall effect, and efforts to understand the high temperature superconductors. In both cases electron interactions are fundamental to the phenomena. However, a new class of systems, non-interacting $Z_2$ topological band insulators (TBI) with time-reversal symmetry (TRS), has diverted attention to topological properties that do not depend on interactions (but are robust to weak interactions).\cite{Kane:prl05,Bernevig:prl06} The existence of topological properties in models that can be exactly solved in the non-interacting limit, and treated to a high degree of accuracy by conventional band theory methods in the interacting limit, has led to precise predictions for experiment.\cite{Fu:prb07,Teo:prb08,Bernevig:sci06,Zhang:np09} An unusually rapid verification of many of these predictions in experiment has followed, and there are now several known examples of this state of matter in both two dimensional \cite{Konig:sci07,Roth:sci09} and three dimensional systems\cite{Hsieh:nat08,Xia:np09,Chen:sci09}. In some of these materials, topological properties are expected to be robust up to room temperature and therefore hold great promise as components of future electronic devices.\cite{Chen:sci09}

Current theoretical research on topological insulators is proceeding along several parallel tracks. On the one hand, there is great interest in identifying new physical systems that will possess topologically non-trivial phases\cite{Guo:prb09,Guo:prl09,Wu:prb09,Schnyder:prl09,Raghu:prl08,Zhang:prb09,Grover:prl08}, while on the other hand there are fundamental questions about the fate of topological properties as the strong electron interaction limit is approached\cite{Young:prb08,Pesin09}. In this work, we contribute to both directions by providing several other examples of $Z_2$ TBIs on a lattice where they have not been reported before--the decorated honeycomb lattice. We also establish a topological connection at 1/2 filling between the non-interacting limit and the strongly interacting limit where an exactly solvable electron model (the Kitaev spin model) is realized on the same lattice.\cite{Yao:prl07} We are unaware of any other model that realizes exactly solvable states at weak and strong interaction, both with topological properties. Moreover, via explicit calculation, we show these two limits share topological properties, even though their symmetries are very different.

\begin{figure}[b]
\includegraphics[width=0.95\linewidth,clip=]{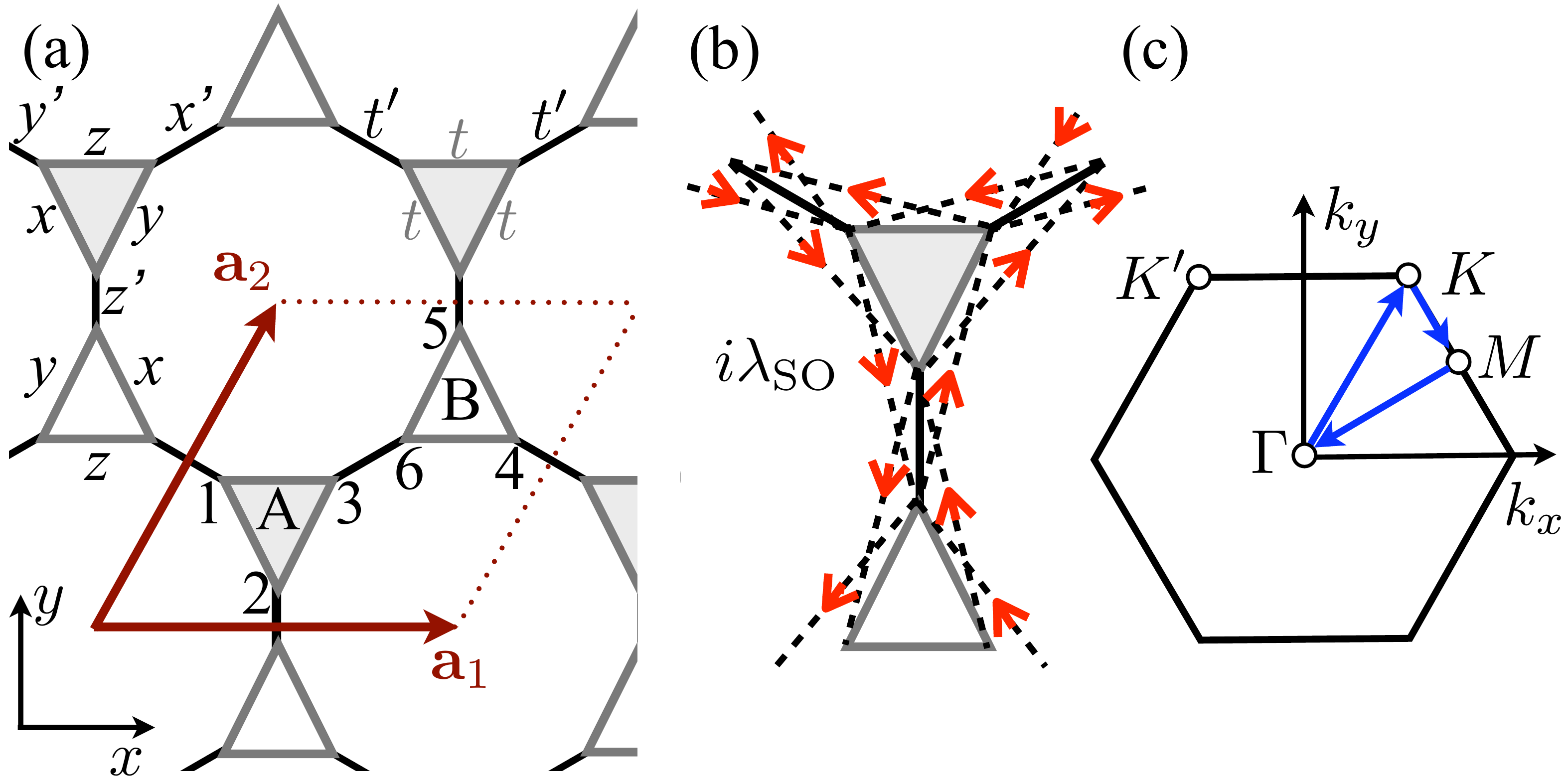}
 \caption{\label{fig:lattice} (color online) (a) The decorated honeycomb lattice has a triangle at each vertex of the honeycomb lattice. The 6-site unit cell with ``sublattice" A and B is contained in the parallelogram indicated by {\bf a}$_1$ and {\bf a}$_2$. Nearest neighbor hopping on vertex triangles occurs with amplitude $t$, between triangles with amplitude $t'$, and with $\pm i\lambda_{\rm SO}$ for second neighbor hopping as indicated in (b). Topological phases occur at a number of filling fractions (see Fig.~\ref{fig:phase_diagram}), as well as in the case that $\lambda_{\rm SO} \equiv 0$ and $t$ is allowed to be complex corresponding to finite flux through vertex triangles. (c) The first Brillouin zone including a path along the high-symmetry lines.}
 \end{figure}

Our discussion focuses on a tight-binding model of fermions hopping on the 2-dimensional decorated honeycomb lattice shown in Fig.~\ref{fig:lattice}. This lattice is a ``cousin" of both the honeycomb lattice and the kagome lattice, each of which is known to support TBI phases. \cite{Haldane:prl88,Ohgushi:prb00,Lee:prl07,Kane:prl05,Guo:prb09} In a certain regard, the decorated honeycomb lattice can be viewed as an ``interpolating" lattice between the honeycomb and the kagome: If one shrinks the triangles at the verticies of the {\em underlying} honeycomb lattice (sites with hopping parameter $t$ in the figure) to their center point, the honeycomb lattice is recovered, while expanding the triangles until their corners touch produces the kagome lattice. One might consider this geometrical property to be the key reason the decorated honeycomb lattice supports topological insulator phases, given that the honeycomb and kagome lattices also support topological phases.  However, because the unit cell of the decorated honeycomb lattice contains 6 sites (compared to 3 for the kagome and 2 for the honeycomb) its phase diagram is much richer than that of either of its ``cousins"  and some novel features appear that we will discuss in more detail below.
\section{Tight-binding model}
The Hamiltonian for our (initially non-interacting) problem is
\begin{equation}
H=H_0+H_{\rm SO}+H_{\rm CDW}+H_{\rm R}.
\label{eq:H}
\end{equation}
The nearest-neighbor hopping is described by
\begin{equation}
\label{eq:H_0}
H_0=-t\!\! \sum_{\langle i j\rangle,\sigma,\Delta} c^\dagger_{i\sigma}c_{j\sigma}
-t' \!\!\!\!\!  \sum_{\langle i j\rangle,\sigma,\Delta \to \Delta} \!\!\!\!\! c^\dagger_{i\sigma}c_{j\sigma}+{\rm h.c.}\\
\end{equation}
with amplitude $t$ on the triangles ``$\Delta$" and with amplitude $t'$ between triangles ``$\Delta \to\Delta$", as shown in Fig.~\ref{fig:lattice}(a). The intrinsic spin-orbit coupling,
\begin{equation}
\label{eq:H_SO}
H_{\rm SO}=i\lambda_{\rm SO}\!\!\! \sum_{\langle \langle i j\rangle \rangle, \alpha, \beta} \!\!\! \vec e_{ij}\cdot {\vec s}_{\alpha \beta}  c^\dagger_{i\alpha}c_{j\beta}+{\rm h.c.},
\end{equation}
describes the second-neighbor hopping with amplitude $\pm i\lambda_{\rm SO}$, see Fig.~\ref{fig:lattice}(b). The sign of the amplitude is different for different spin orientations $s_z=\pm 1$, $\vec s$ is the vector of Pauli matricies and $\vec e_{ij}=({\bf d}^1_{ij}\times {\bf d}^2_{ij})/|{\bf d}^1_{ij}\times {\bf d}^2_{ij}|$ is a vector normal to the $x-y$ plane describing how the path $\langle \langle i j\rangle \rangle$ was traversed using the standard conventions.\cite{Kane:prl05} In Eq.~\eqref{eq:H}, $H_{\rm CDW}$ and $H_{\rm R}$ are charge density wave and Rashba spin-orbit terms, respectively. The CDW Hamiltonian is
\begin{equation}
\label{eq:H_CDW}
H_{\rm CDW}=\sum_{i,\sigma} \lambda_{vi} c^\dagger_{i\sigma}c_{i\sigma},
\end{equation}
and the Rashba Hamiltonian is
\begin{equation}
\label{eq:H_R}
H_{\rm R}=i\lambda_{\rm R} \sum_{\langle i j\rangle,\alpha,\beta} c^\dagger_{i\alpha}(\vec s_{\alpha \beta} \times \hat d_{ij})_zc_{j\beta}+{\rm h.c.},
\end{equation}
where $\lambda_{vi}$ is an on-site potential possibly differing on each of the 6 unit cell sites labeled in Fig.~\ref{fig:lattice}, $\lambda_{\rm R}$ is the strength of the Rashba coupling and $\hat d_{ij}$ is the unit vector connecting site $i$ to $j$.

\begin{figure}[h]
\includegraphics[width=0.95\linewidth,clip=]{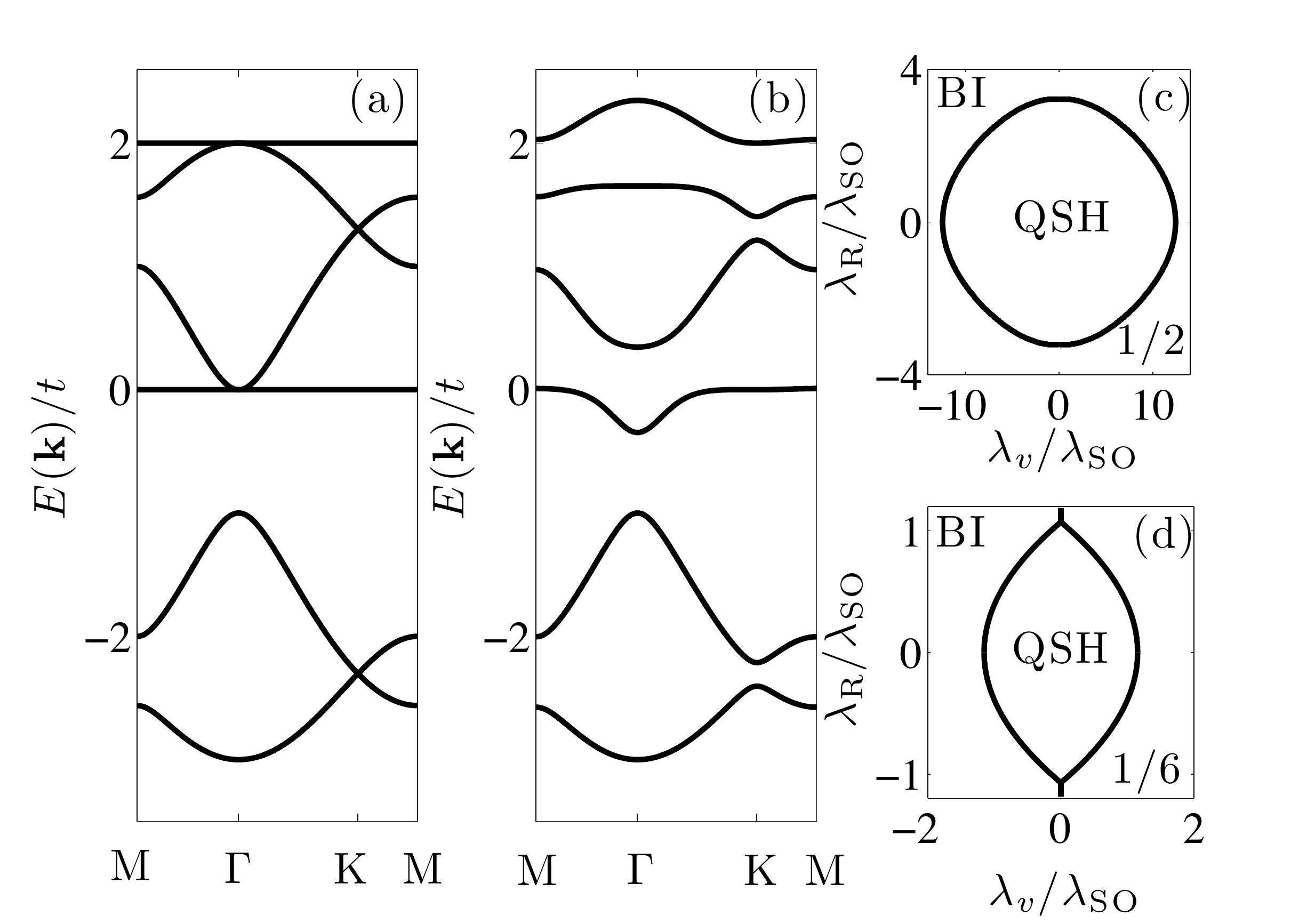}
\caption{\label{fig:band_structure} (a) and (b) show the band structure of the tight binding model $H_0+H_{\rm SO}$ with $t=t'$ along the path shown in Fig.~\ref{fig:lattice}(c). In (a) $\lambda_{\rm SO}=0$ and in (b) $\lambda_{\rm SO}=0.1t$.  There are Dirac points at $K$ and $K'$ (not shown) and quadratic band crossing points (QBCP) at $\Gamma$ in (a), while in (b) $\lambda_{\rm SO}\neq 0$ opens up a gap at each of these points and destroys the flat bands. (c) The phase diagram at $f=1/2$ (involving QBCP) with $\lambda_{\rm SO}=0.1t$. (d) The phase diagram at $f=1/6$ (involving Dirac points) for $\lambda_{\rm SO}=0.1t$. We have chosen a staggered sublattice potential configuration where all the sites in A(B)-triangle (see Fig.~\ref{fig:lattice}) have potentials $\lambda_{v}$ ($-\lambda_{v}$).}
\end{figure}

\section{Phase diagrams}

\begin{figure}[h]
\includegraphics[width=.95\linewidth,clip=]{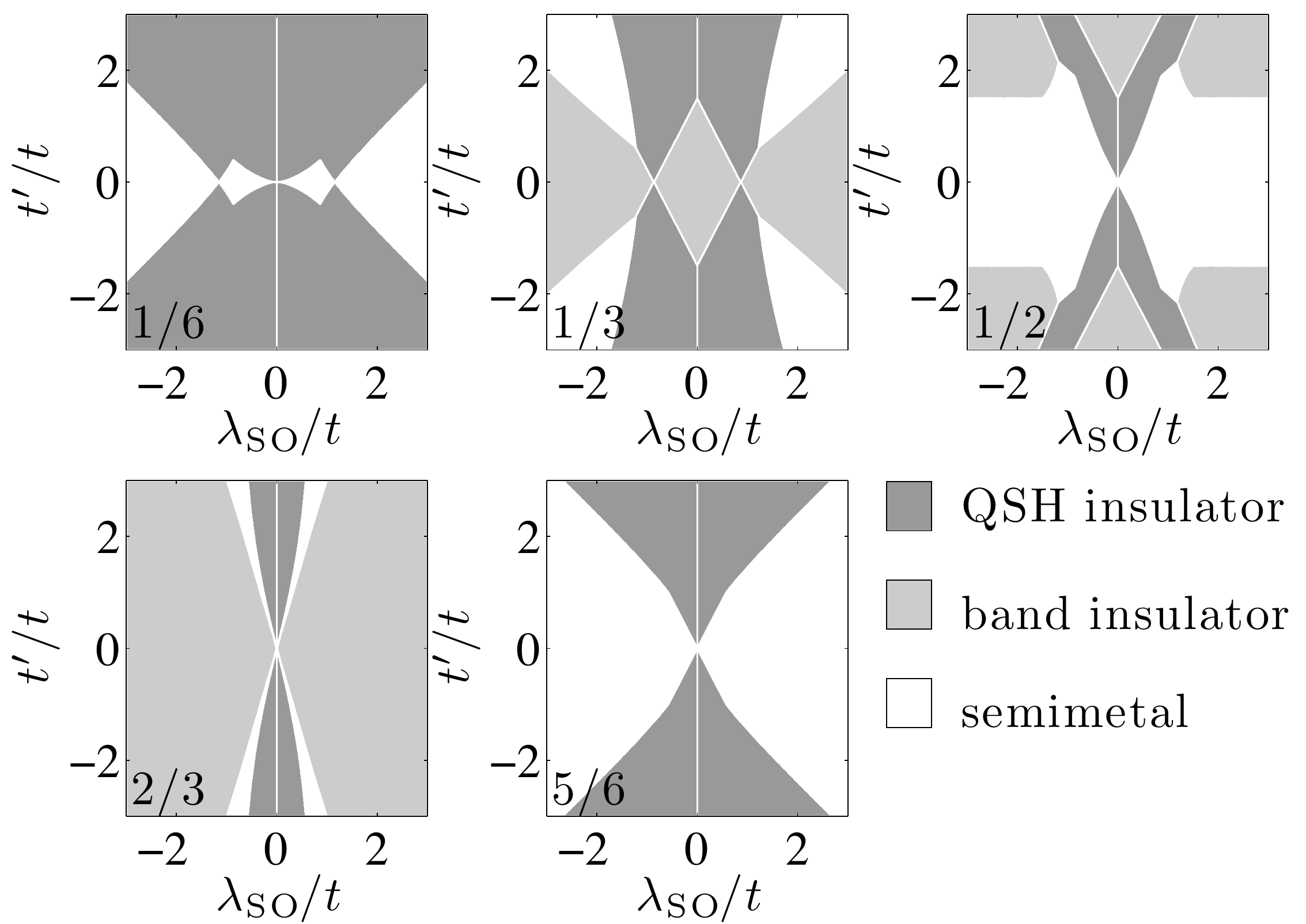}
\caption{\label{fig:phase_diagram}  Phase diagrams for the decorated honeycomb lattice with $t$ and $t'$ real in the absence of a staggered on-site potential and no Rashba coupling. Several filling fractions $f$ are shown (lower left corner). For fixed $f$ and $\lambda_{\rm SO}$ it is possible to  drive a transition between a topological insulator and a non-topological phase by varying the ratio $t'/t$.}
\end{figure}

The 6 (doubly degenerate) bands coming from the 6-site unit cell (see Fig.~\ref{fig:lattice}) for $H_0+H_{\rm SO}$ are shown in Fig.~\ref{fig:band_structure} along the various high symmetry directions.  The first   Brillouin zone is identical to that of the honeycomb and kagome lattices which share the same underlying triangular Bravais lattice,\cite{Kane:prl05,Guo:prb09} see Fig.~\ref{fig:lattice}(c). There are Dirac points at $K$ and $K'$, two quadratic band crossing points (QBCP) at $\Gamma$, and two flat bands present when $\lambda_{\rm SO}=0$. We note that the lower QBCP appears at filling fraction $f=1/2$ for $t'<3t/2$ and at $f=1/3$ for $t'>3t/2$. Similar band features are also found on the kagome lattice at the same Brillouin zone points.\cite{Guo:prb09,Ohgushi:prb00} When the second neighbor hopping $\lambda_{\rm SO} \neq 0$, a gap opens at the Dirac and the QBCP and topologically non-trivial phases appear; denoted as quantum spin Hall (QSH) insulator in Figs.~\ref{fig:band_structure} and \ref{fig:phase_diagram}. By explicitly computing the $Z_2$ invariant using the parity eigenvalues at the time-reversal invariant moment\cite{Fu:prb07} and checking for helical edge states in a strip geometry\cite{Kane:prl05}, we have found the phase diagrams for different filling fractions, $f$. The results are summarized in Fig.~\ref{fig:phase_diagram}.

One feature of the decorated honeycomb lattice that differs from the kagome and honeycomb lattices is the natural presence of two ($t$ and $t'$), rather than one, nearest-neighbor hopping parameters. This effectively adds an additional degree of freedom to the phase diagram and can lead to transitions to topologically non-trivial phases even when there is not an {\em obvious} Dirac point or quadratic band crossing involved in the nearest-neighbor hopping model, such as occur at filling fraction 1/3 in Fig.~\ref{fig:band_structure}(a). As Fig.~\ref{fig:band_structure}(b) shows, when $\lambda_{\rm SO}$ is turned on, an ``incipient" band touching point develops at the $\Gamma$ point for filling fraction 1/3 and this effectively drives the transition to the topologically non-trivial state. Thus, the band structures with zero spin-orbit coupling do not always clearly reveal potential topological transitions for strong spin-orbit coupling.

At $f=1/3, 1/2$, and $2/3$ there are (electron-hole compensated) intervening metallic phases between topologically trivial and non-trivial insulators. This also indicates that a ``direct" transition coming from a band inversion is not generic in this model.\cite{Murakami:prb07} The filling fractions with such an intervening metallic phase mimic the behavior of disorder on the honeycomb lattice in the presence of finite Rashba coupling.\cite{Essin:prb07} Also note that varying the ratio of $t'/t$ at fixed $\lambda_{\rm SO}$ can lead to a transition between a TBI and a trivial insulator. More surprisingly, increasing $\lambda_{\rm SO}$ for fixed value of $t'/t$ can trigger a transition between a TBI and a trivial insulator, as seen for $f=1/3$ and $1/2$.

Next we turn to an analysis of the stability of the topological phases indicated in Fig.~\ref{fig:phase_diagram} in the presence of Rashba interaction and on-site (CDW) potentials. The stability of the topological phases at $f=1/6$ and $f=1/2$ is shown in Fig.~\ref{fig:band_structure} (c) and (d), where we used $\lambda_{vi}=\lambda_v$ on ``sublattice" A and $\lambda_{vi}=-\lambda_v$ on ``sublattice" B as shown in Fig.~\ref{fig:lattice}. Consequently, the stability regions are qualitatively similar to the analagous model on the honeycomb lattice.\cite{Kane:prl05,Sheng:prl06} The stability region of the QBCP at $f\!=\!1/2$ is larger than that for the Dirac point at $f\!=\!1/6$, which we attribute to a larger value of the gap ($\sim$3 times) at the QBCP when $\lambda_R=\lambda_v=0$.\cite{Sun:prl09}

\section{Effective low energy description at Dirac points}
At the Dirac points it is straightforward to derive an effective low energy description for arbitrary $\lambda_{vi}$ in the 6-site unit cell.  Setting the zero of energy to be right at the Dirac point (either at $f=1/6$ or $f=2/3$) for $\lambda_{\rm SO}=\lambda_R=\lambda_{vi}=0$, the effective low-energy Hamiltonian is given by 
\begin{equation}
H'=H'_{0}+H'_{\rm SO}+H'_{\rm R}+H'_{\rm CDW}
\end{equation}
with
\begin{eqnarray}
H'_{0}&=&\alpha v_{F}\hbar (k_{x}\tau _{z}\sigma _{x}+k_{y}\sigma _{y}),\label{eq:H0p}\\
H'_{\rm SO}&=&-4\alpha\, w(t'/t)\,\lambda _{\rm SO}\sigma _{z}\tau_{z}s_{z},\\ 
H'_{\rm R}&=&-\alpha\, w_{R}(t'/t)\,\lambda _{R}(\sigma _{x}\tau_{z}s_{y}-\sigma _{y}s_{x}),\label{eq:HSOp}\\
H'_{\rm CDW}&=&g_{0}I+\alpha (g_{x}\sigma _{x}+g_{y}\tau _{z}\sigma _{y}+g_{z}\sigma_{z}).\label{eq:HCDWp}
\end{eqnarray}
We have adopted a $\tau_z, \sigma_i,s_i$ notation similar to Ref.~[\onlinecite{Kane_2:prl05}]: The $\tau_z=\pm 1$ describes states at either the $K$ or $K'$ points, the $\sigma_z=\pm 1$ describes the two bands that are involved in the Dirac band crossing (analog of A and B sublattice bands on the honeycomb lattice), and $s_z=\pm 1$ represents the electron spin as it did in Eq.~\eqref{eq:H_SO} and Eq.~\eqref{eq:H_R}. The parameter $\alpha =\pm 1$ refers to the Dirac point at $f=1/6$, and $f=2/3$, respectively. We have also defined two functions describing the dependence of the effective low-energy theory on the parameter $x=t'/t$:
\begin{eqnarray*}
w(x)&=&\frac{\sqrt{3}|x|}{2\sqrt{9+4x^{2}}},\\
w_{R}(x)&=&\frac{3+2\sqrt{3}x}{\sqrt{9+4x^{2}}}.
\end{eqnarray*}
The effective Fermi velocity entering Eq.~\eqref{eq:H0p} is
\begin{equation*}
v_{F}=w(t'/t)v_0
\end{equation*}
where $v_{0}=ta/\hbar$ and $a$ is the length of the unit cell vector. It follows from Eq.~\eqref{eq:HSOp} that for $\lambda_{vi}=\lambda_{\rm R}=0$ the spin-orbit coupling opens up a gap with magnitude $E_{\rm gap}=8|w(t'/t)\lambda _{\rm SO}|$.
The parameters entering the low-energy description of the CDW term, Eq.~\eqref{eq:HCDWp}, are given by
\begin{eqnarray*}
g_{0} &=&\frac{\lambda _{v 1}\!+\!\lambda _{v 2}\!+\!\lambda _{v 3}\!+\!\lambda
_{v 4}\!+\!\lambda _{v 5}\!+\!\lambda _{v 6}}{6},\\
g_{x} &=&w\!\left(t^{\prime }/t\right)\frac{\lambda _{v 1}\!+\!\lambda _{v 2}\!-\!2\lambda _{v 3}\!+\!\lambda _{v 4}\!+\!\lambda _{v 5}\!-\!2\lambda
_{v 6}}{3\sqrt{3}},\\
g_{y}&=&w\!\left(t'/t\right)\frac{\!-\!\lambda _{v 1}\!+\!\lambda _{v 2}\!-\!\lambda _{v 4}\!+\!\lambda
_{v 5}}{3},\\
g_{z} &=&w\!\left(t'/t\right)\frac{-\!\lambda _{v 1}\!-\!\lambda _{v 2}\!-\!\lambda _{v 3}\!+\!\lambda
_{v 4}\!+\!\lambda _{v 5}\!+\!\lambda _{v 6}}{\sqrt{3}(t^{\prime }/t)}.
\end{eqnarray*}
A finite $g_0$ can be absorbed in a shift of the chemical potential.

It is useful to consider a few important limits of the general low-energy form of $H_{\rm CDW}$. First take $\lambda_{vi}\!=\!\lambda_v$ for sites on the A-triangle and $\lambda_{vi}\!=\!-\lambda_v$ for sites on the B-triangle.  In this case, 
\begin{equation*}
H_{\rm CDW}\!=\!-\alpha \frac{2\sqrt{3}w(\frac{t^{\prime }}{t})\lambda _{v }}{t^{\prime }/t}\sigma _{z},
\end{equation*}
which is identical to the form of the expression for the honeycomb lattice and will generically open a gap at the Dirac point.\cite{Kane:prl05} We have verified that the low-energy description given above produces the same stability phase diagram and phase boundary shown in Fig.~\ref{fig:band_structure}(d) as a direct diagonalization of the full 6-band Hamiltonian. Another important limit to consider is that of general $\lambda_{vi}$.  In that case, the physics more closely resembles the kagome lattice where an effective axial gauge field appears \cite{Guo:prb09}
with 
\begin{eqnarray*}
{\cal A}^l_x&=&-\frac{\lambda _{v 1}\!+\!\lambda _{v 2}\!-\!2\lambda _{v
3}\!+\!\lambda _{v 4}\!+\!\lambda _{v 5}\!-\!2\lambda _{v 6}}{3\sqrt{3}}l,\\ 
{\cal A}^l_y&=&\frac{\lambda _{v 1}-\lambda _{v 2}+\lambda _{v
4}-\lambda _{v 5}}{3}l
\end{eqnarray*}
when $\lambda _{v 1}\!+\!\lambda _{v 2}\!+\!\lambda _{v3}\!-\!\lambda _{v 4}\!-\!\lambda _{v 5}\!-\!\lambda _{v 6}\!=\!0$, where $l=\pm 1$ refers to the two Dirac points $K$ and $K'$.  If, on the other hand, $\lambda _{v 1}\!+\!\lambda _{v 2}\!+\!\lambda _{v3}\!-\!\lambda _{v 4}\!-\!\lambda _{v 5}\!-\!\lambda _{v 6}\neq 0$ a gap 
\begin{equation*}
E_{\rm gap}=\frac{|\lambda _{\nu 1}+\lambda _{\nu 2}+\lambda _{\nu
3}-\lambda _{\nu 4}-\lambda _{\nu 5}-\lambda _{\nu 6}|}{\sqrt{9+4(t^{\prime
}/t)^{2}}}
\end{equation*}
opens with a smallest direct gap at shift ${\cal A}^l_x, {\cal A}^l_y$ with respect to $K$ or $K'$. 
Thus, the behavior of the decorated honeycomb lattice with respect to $H_{\rm CDW}$ is another example of the ways in which this lattice ``interpolates" between the honeycomb and kagome lattices, and we expect, for example, analogs of the kekule phase to be realized as well.\cite{Guo:prb09,Hou:prl07}
\section{Adiabatic deformations and the Kitaev model}
\begin{figure}[h]
\includegraphics[width=0.9\linewidth,clip=]{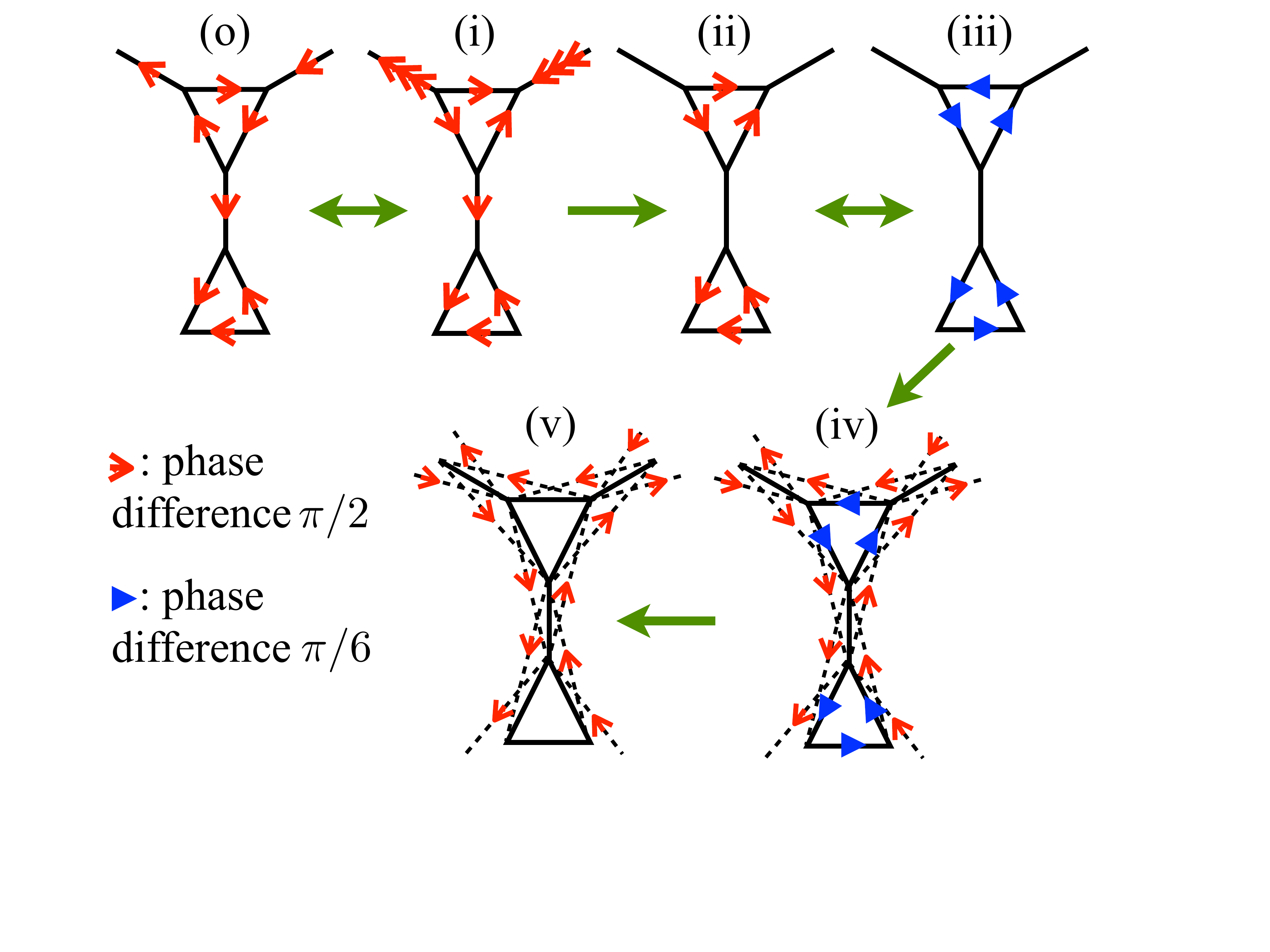}
\caption{\label{fig:path}(color online) Schematic illustration of the continuous path which adiabatically connects the model (o) and (i) lacking second neighbor hopping but having complex $t$ (the representative free fermion model of the ground state sector of the Kitaev model on the decorated Honeycomb lattice) with the spinless model (v) with real $t$, $t'$ and $\lambda_{\rm SO}$, denoted by (v). Along this path, the continuous deformation does not lead to a gap closing and the Chern number stays constant. This establishes the topological connection between the two models.}
\end{figure}

We now turn our attention to one of the features of the decorated honeycomb lattice which is related to its geometry: Topological phases exist {\em even in the absence of second neighbor hopping} when $t$ is made complex (obtained by putting a flux through the vertex triangles). Below we show by an explicit calculation for spinless fermions that the model obtained in the absence of second neighbor hopping but with complex $t$ (and possibly also complex $t'$) can be adiabatically deformed into a model with real $t$, $t'$ and $\lambda_{\rm SO}$. An example of such an adiabatic deformation is illustrated in Fig.~\ref{fig:path} and in the last part of this section we will describe each step of the deformation in detail.

The adiabatic connection we establish also holds for time reversal invariant models of electrons with spin: For $s_z$ conserving models on the honeycomb lattice, Kane and Mele showed \cite{Kane:prl05,Kane_2:prl05} that one can view a $Z_2$ TBI in 2-d as two copies of Haldane's model \cite{Haldane:prl88} with different effective magnetic fluxes (with a net zero flux through the unit cell) for different spins (so that under time-reversal each copy is transformed into the other, thus preserving TRS overall). Moreover, as long as the gap does not close, $s_z$ non-conserving terms are also allowed. With this insight, it is evident that any lattice model that supports a quantum Hall effect for spinless fermions will support a $Z_2$ TBI for electrons with spin (by taking the appropriate ``second copy").

There is an interesting consequence of the above mentioned adiabatic connection. It allows us to topologically relate the phases in the non-interacting tight-binding model at half-filling, see Fig.~\ref{fig:phase_diagram}, to the chiral spin liquid phases recently reported in the Kitaev model\cite{Kitaev:2006, Yao:prl07} on the same lattice which can be viewed as a {\em strongly interacting} electron model with spin-orbit coupling\cite{Jackeli:prl2009}. The Kitaev model on the decorated honeycomb lattice is defined in the following way\cite{Yao:prl07, Dusuel:prb2008}
\begin{eqnarray*}\label{eq_model}
{\cal H}&=&\sum_{x\textrm{-link}} J_x\sigma^x_i\sigma^x_j +\sum_{y\textrm{-link}}J_y\sigma^y_i\sigma^y_j +\sum_{z\textrm{-link}}J_z\sigma^z_i\sigma^z_j\nonumber\\ 
&&+\sum_{x'\textrm{-link}} J'_{x}\sigma^x_i\sigma^x_j +\sum_{y'\textrm{-link}}J'_{y}\sigma^y_i\sigma^y_j +\sum_{z'\textrm{-link}}J'_{z}\sigma^z_i\sigma^z_j.
\end{eqnarray*}
In the summation, $i$ and $j$  are nearest neighboring sites connected by a $\alpha$-link ($\alpha=x,y,z,x',y',z'$) as shown in Fig.~\ref{fig:lattice}(a). After a Jordan-Wigner transformation this model can be mapped onto a model of free majorana fermions hopping in the background of static $Z_2$ fluxes. The ground state spontaneously breaks time-reversal symmetry and is described by a chiral spin liquid with either Abelian or non-Abelian vortex excitations.\cite{Yao:prl07} In the following we set $J_x=J_y=J_z=J>0$ and $J_x'=J_y'=J_z'=J'>0$. The ground-state sector corresponds to a uniform flux configuration. It possesses a representative free fermion model\cite{Lee:prl07}
\begin{equation}
H_{\rm CSL}=\gamma \sum_{{\bf k}}\psi_{{\bf k}\alpha}^{\dag}H_{\alpha\beta}^{\rm (o)}({\bf k})\psi_{{\bf k}\beta}
\label{eq:HCSL}
\end{equation}
and $\gamma=\pm1$ specifies the way the time reversal symmetry is spontaneously broken.
Here, $\psi_{{\bf k}\alpha}^{(\dag)}$, $\alpha=1,\dots,6$ are fermionic annihilation (creation) operators and we have defined the matrix
\begin{eqnarray*}
&&H^{\rm (o)}=\\
&&\begin{pmatrix}
0 & iJ & iJ& -iJ'e^{-ik_2}&0&0\\
-iJ &0&-iJ&0&iJ'e^{ik_1}&0\\
-iJ&iJ&0&0&0&iJ'\\
iJ'e^{ik_2}&0&0&0&iJ&-iJ\\
0&-iJ'e^{-ik_1}&0&-iJ&0&iJ\\
0&0&-iJ'&iJ&-iJ&0
\end{pmatrix}.
\end{eqnarray*}
The flux pattern derived from the matrix $H^{\rm (o)}$ is illustrated in (o) of Fig.~\ref{fig:path}. The spectrum of Eq.~\eqref{eq:HCSL} is gapped at half filling as long as $J'\neq\sqrt{3}J$ and the Chern number is $\nu=-\gamma$ for $J'<\sqrt{3}J$ and $\nu=0$ for $J'>\sqrt{3}J$.\cite{Yao:prl07} The two sectors connected by time reversal symmetry and characterized by the parameter $\gamma=\pm 1$ are similar to the two ``copies" characterized by $s_z=\pm1$ in the Kane-Mele type model. However, in the strongly interacting limit (Kitaev model) the system spontaneously chooses one sector whereas the non-interacting TBI model involves a summation over the two sectors (spin).

\begin{figure}
\centering
\includegraphics[width=0.85\linewidth]{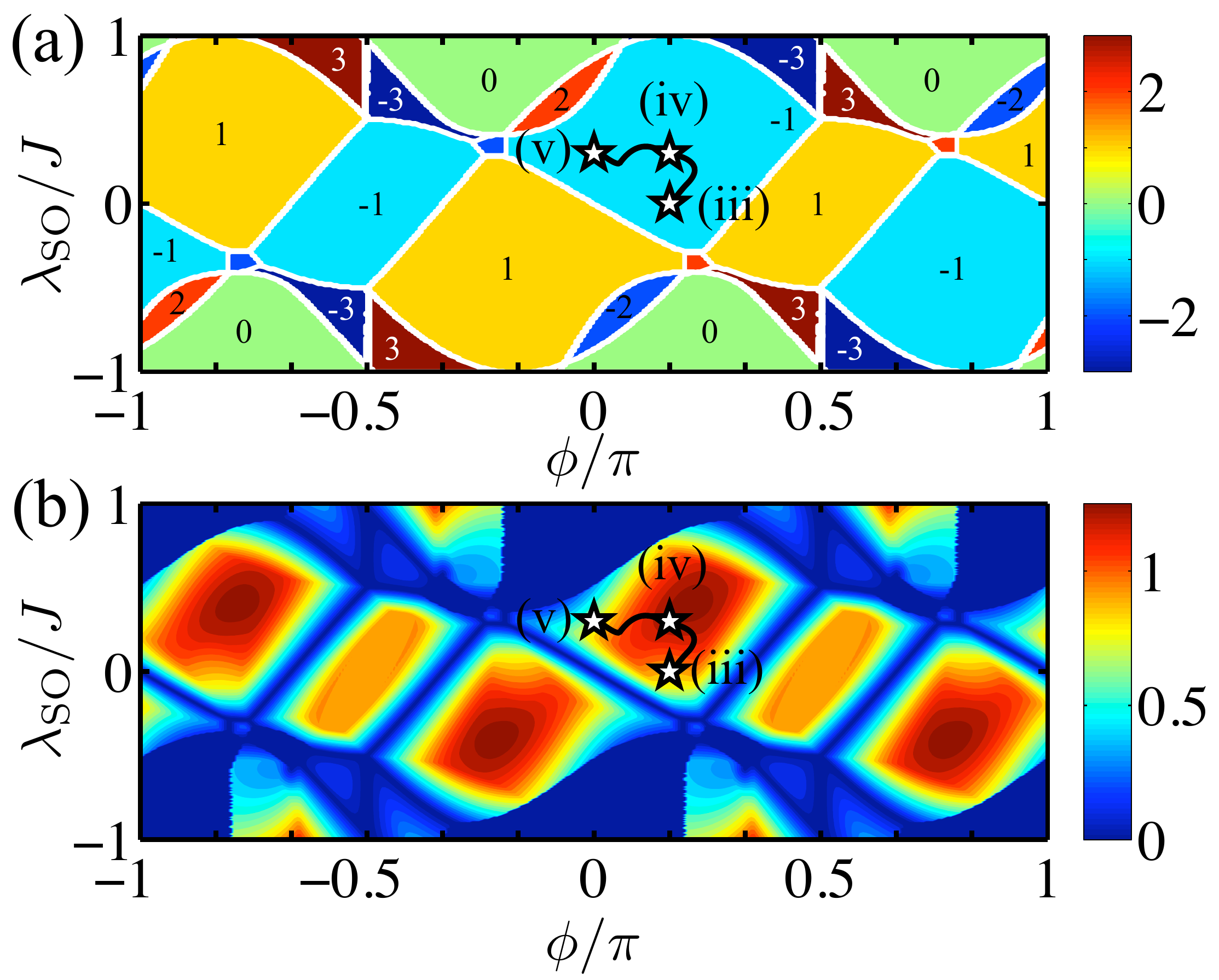}
\caption{(color online) Contour plot of (a) the Cern number and (b) the gap of the model defined by $\Gamma(\phi,\lambda_{\rm SO})$ at half filling for $J=J'$. Also shown is a possible path which adiabatically connects the flux patterns (iii), (iv) and (v) defined in Fig.~\ref{fig:path}.}
\label{fig:iiitov}
\end{figure}

We now discuss the adiabatic connection of the model $H^{\rm (o)}$ to the spinless model with real $t$, $t'$ and $\lambda_{\rm SO}$; model $H^{\rm (v)}$ in Fig.~\ref{fig:path}. As the starting point for the continuous deformation we use a gauge equivalent pattern (i) obtained from (o) by replacing $\psi_{{\bf k}4}\rightarrow-\psi_{{\bf k}4}$ and $\psi_{{\bf k}5}\rightarrow-\psi_{{\bf k}5}$. From (i) to (ii) the phase difference for hopping between the triangles is gradually reduced to zero. This can be achieved by replacing $\pm iJ'$ in $H^{\rm (i)}$ by $\exp(\pm is\pi)J'$ and continuously reducing $s$ from 1 to 0. This process does not change the fluxes through the triangles and the dodecagons; instead the global fluxes are modified and the whole spectrum is moved in ${\bf k}$-space according to $k_1\rightarrow k_1+(1-s)\pi$ and $k_2\rightarrow k_2$. Consequently, the direct gap stays constant. From (ii) to (iii) a continuous gauge transformation is applied to change the phase difference for hopping within the triangles from $\pm\pi/2$ to $\pi/6$ (this does not modify any flux and the gap remains constant). From (iii) to (v) we turn on $\lambda_{\rm SO}$ as shown in (iv) and make $t$ real. Along this path the value of the gap varies in general. 
Explicitly, we can define the set of matrices
\begin{eqnarray*}
&&\Gamma(\phi,\lambda_{\rm SO})=\\
&&\begin{pmatrix}
0 & e^{-i\phi}J &e^{i\phi}J& J'e^{-ik_2}&D&-E\\
e^{i\phi}J &0&e^{-i\phi}J&-D&J'e^{ik_1}&F\\
e^{-i\phi}J&e^{i\phi}J&0&E&-F&J'\\
J'e^{ik_2}&-D^*&E^*&0&e^{-i\phi}J&e^{i\phi}J\\
D^*&J'e^{-ik_1}&-F^*&e^{i\phi}J&0&e^{-i\phi}J\\
-E^*&F^*&J'&e^{-i\phi}J&e^{i\phi}J&0
\end{pmatrix}.
\end{eqnarray*}
Here we have introduced
\begin{eqnarray*}
D&=&-i\lambda_{\rm SO}\left(e^{-ik_1}+e^{-ik_2}\right),\\
E&=&-i\lambda_{\rm SO}\left(1+e^{-ik_1}\right),\\
F&=&-i\lambda_{\rm SO}\left(1+e^{-ik_2}\right).
\end{eqnarray*}
Clearly, $\Gamma(\pi/6,0)=H^{\rm (iii)}$ and $\Gamma(0,\lambda_{\rm SO})=H^{\rm (v)}$. By numerical examination of the gap and of the Chern number\cite{Fukui:jpsj05} at half-filling we find that there is a large range of parameters which allows one to adiabatically connect the model (iii) with the model (v). This is shown in Fig.~\ref{fig:iiitov} where we plot the Chern number in (a) and the value of the gap in (b) obtained for $J=J'$. 

The model defined by $\Gamma(\phi,\lambda_{\rm SO})$ shows a complex phase diagram with a variety of topological phases distinguished by different values of the Chern number. As long as there is a direct gap, the Chern number is well-defined and regions with different values are necessarily separated by gap closings. However, there are also regions in parameter space where an indirect gap is closed indicating the presence of partially filled bands at half-filling, see Fig.~\ref{fig:iiitov}(b). The phase diagram of $\Gamma(\phi,\lambda_{\rm SO})$ also depends on the ratio $J'/J$. For $J'>\sqrt{3}J$ the Chern number of $H^{\rm (iii)}$ is zero and the connection (iii) to (v) holds between topologically trivial phases. 

These arguments demonstrate the adiabatic connection between the strongly interacting chiral spin liquid phases of the Kitaev model and the phases obtained from the spinless model at half-filling with real $t$, $t'$ and $\lambda_{\rm SO}$. In this sense, we also establish a connection to the $Z_2$ TBI when two ``copies" of the spinless model are taken. It can be shown that a generalized spin-3/2 Kitaev model on the kagome lattice also supports a chiral spin liquid ground state with non-Abelian excitations \cite{Yao} and similar arguments can be made for connections to other topological phases.\cite{Fiete}

\section{Summary}
In summary, we have shown that the decorated honeycomb lattice supports $Z_2$ topological phases at various filling fractions, discussed their stability, and described the similarities and differences with the $Z_2$ topological phases on the kagome\cite{Guo:prb09} and honeycomb lattices\cite{Kane:prl05}. The limit of weak spin-orbit coupling yields phase diagrams which are very similar to the ones observed on the honeycomb or kagome lattice. This observation is in agreement with the conclusions drawn from the effective low-energy theory at the Dirac points. The situation for strong spin-orbit coupling can be quite different and leads to novel aspects. One surprising observation is that a large $\lambda_{\rm SO}$ can trigger a transition from a TBI to a trivial insulator. 

We have also shown that the tight-binding models with real $t$, $t'$ and $\lambda_{\rm SO}$ can be adiabatically connected to models without second-neighbor hopping but with complex $t$ (and possibly also complex $t'$). This property was explicitly demonstrated at half filling by a calculation of the gap and the Chern number. Moreover, we have argued that this adiabatic connection allows us to topologically relate the chiral spin liquid phases recently discovered on the decorated honeycomb lattice to the phases obtained at half-filling in the non-interacting TBI model. Our work therefore provides an example of a non-interacting and a strongly interacting model defined on the same lattice which are both exactly solvable and show topologically related states. To determine the precise form of the spin-orbit interaction, the number, and the size (coupling strength) of the terms needed in a generalized extended Hubbard model at half filling to interpolate between the topological band insulator and the Kitaev model on the decorated honeycomb lattice is an interesting open problem beyond the scope of this work. However, based on our results here and related studies that realize Kitaev models in certain low energy limits\cite{Jackeli:prl2009,Wang} we believe that such an interacting microscopic model can be found. 
 
Finally, we note that an underlying ``star" lattice has been experimentally reported for a polymeric Iron(III) acetate\cite{Zheng:acie07}, and some of our results may be relevant for this solid state example of a decorated honeycomb lattice. We also believe it is possible to realize much of the discussed physics (including Kitaev models), in cold atomic gases, given that its two cousins, the honeycomb and kagome lattices, can be realized in optical lattices.\cite{Duan:prl2003,Ruotoski:prl2009}

\acknowledgments
We gratefully acknowledge support from ARO grant  W911NF-09-1-0527.

$^*$jwen@physics.utexas.edu

\end{document}